\documentclass[%
 aip,
 amsmath,
 amssymb,
 reprint,
]{revtex4-1}

\usepackage{graphicx}
\usepackage{dcolumn}
\usepackage{bm}

\usepackage{url}
\usepackage[utf8]{inputenc}
\usepackage[T1]{fontenc}
\usepackage{mathptmx}
\usepackage{etoolbox}
\usepackage{amsmath}
\usepackage{hyperref}
\usepackage{dcolumn}

\makeatletter
\def\@email#1#2{%
 \endgroup
 \patchcmd{\titleblock@produce}
  {\frontmatter@RRAPformat}
  {\frontmatter@RRAPformat{\produce@RRAP{*#1\href{mailto:#2}{#2}}}\frontmatter@RRAPformat}
  {}{}
}

\makeatother
\begin{document}
\title{Computational modelling of Parkinson's disease: A multiscale approach with deep brain stimulation and stochastic noise} 

\author{A. Herrera*}
 \email[]{herrera8@myumanitoba.ca}
\affiliation{Department of Statistics, University of Manitoba, Winnipeg, Canada}

\author{H. Shaheen}
\affiliation{Department of Statistics, Faculty of Science, University of Manitoba, Winnipeg, Canada}

\date{\today}

\begin{abstract}
Multiscale modelling presents a multifaceted perspective into understanding the mechanisms of the brain and how neurodegenerative disorders like Parkinson's disease (PD) manifest and evolve over time. In this study, we propose a novel co-simulation multiscale approach that unifies both micro- and macroscales to more rigorously capture brain dynamics. The presented design considers the electrodiffusive activity across the brain and in the network defined by the cortex, basal ganglia, and thalamus that is implicated in the mechanics of PD, as well as the contribution of presynaptic inputs in the highlighted regions. The application of deep brain stimulation (DBS) and its effects, along with the inclusion of stochastic noise are also examined. We found that the thalamus exhibits large, fluctuating spiking in both the deterministic and stochastic conditions, suggesting that noise contributes primarily to neural variability, rather than driving the overall spiking activity. Ultimately, this work intends to provide greater insights into the dynamics of PD and the brain which can eventually be converted into clinical use.
\end{abstract}

\maketitle 

\begin{quotation}
Computational modelling has proved to be an invaluable tool in attempts to discern the intricacies of the brain. The results of research in this field can have many different implications, including explanations to the underlying mechanisms of cognition or the origin and development of neurodegenerative diseases (NDDs). This study focuses on Parkinson's disease (PD), a type of NDD that can degrade the quality of life by inducing movement disorders and other non-motor symptoms. We strive to clarify the dynamics of PD and see how deep brain stimulation --- a common treatment given to those with PD and other NDDs --- affects the brain when in the Parkinsonian state. This study accomplishes this by incorporating a model that links the large-scale behaviour of diffusion across the brain, as well as a microscale environment that simulates the neuron-level functioning for regions in the basal ganglia and thalamus which are significant constituents in PD development. The brain is also characterized by randomness (i.e. noise), an innate feature in its processes. Hence, our design also involves a noise term. In general, we found that this model reflects the inherent unpredictability of neural firing within the brain.
\end{quotation}

\section{\label{sec:level1}Introduction}

The basal ganglia (BG) is a group of brain nuclei responsible for various tasks, including executive functions and emotion regulation. However, its primary purpose is for motor control. \cite{lanciego_functional_2012,graybiel_basal_2000} In greater detail, the BG is composed of the caudate nucleus, the lenticular nucleus — which consists of the putamen, globus pallidus externus (GPe), and the globus pallidus internus (GPi) — the subthalamic nucleus (STN), and the substantia nigra (SN). \cite{young_neuroanatomy_2025} Located in the center of the brain, the thalamus (TH) plays a crucial role in passing information to different regions as motor, limbic, and sensory pathways all coincide in this region. \cite{sheridan_neuroanatomy_2025} Thus, the TH also contributes to our movement, emotional control, and the processing of external stimuli through our senses. 

Although the basal ganglia and thalamus (BGTH) individually are imperative to our normal functioning, the higher processes that we are capable of, such as motor movement, are ultimately shaped by the dynamic between these regions and the entire brain network collectively. Previous studies have been able to highlight the significant connections between these regions and produce a detailed circuitry of the excitatory and inhibitory inputs that these structures receive. \cite{smith_editorial_2022,rubin_basal_2012} Despite these strides in unravelling the intricate interactions in the BGTH, gaps in the behaviour of other components and how they relate to one another still exist. For example, the origin and pathology of numerous neurodegenerative disorders (NDDs) that relate heavily to the BGTH, such as Parkinson’s disease, remain unknown. Parkinson’s disease (PD) is one of the most common NDDs (only behind Alzheimer’s disease), affecting over six million people globally.\cite{armstrong_diagnosis_2020} It is characterized by the loss of neurons in the SN. These neurons are essential in the creation of dopamine, a vital part of the body that creates smooth movement. Consequently, various forms of motor impairment can occur, including tremors, muscle rigidity, bradykinesia, and postural instability. \cite{nih_pd,ramesh_depletion_2023} Additionally, individuals with PD have an accumulation of $\alpha$-synuclein which compose Lewy bodies in different brain regions, often affecting cognition. \cite{lees_parkinsons_2009} While the exact origin of PD is largely unknown, it is believed that a combination of genetics and exposure to several environmental components, such as pesticides and neurotoxins, is associated with a higher risk of developing PD. \cite{bloem_parkinsons_2021,kalia_parkinsons_2015} Other factors that complicate our understanding of PD in addition to the unclear etiology, are the inconsistent symptoms and dissimilar pathogenesis among separate individuals who suffer from this disease. With how hindering this disease can be and the fact that no treatment yet exists to slow down its progress, it is critical to address the elements of PD that still lack clarity. Although there is no definitive cure for PD currently, there are several treatments that help alleviate the associated symptoms, one of them being deep brain stimulation (DBS). High-frequency DBS was first used in 1997. \cite{benabid_deep_2003} It was intended to be a safer alternative to thalamotomies --- another type of treatment where a lesion would be made in the thalamus to prevent the movement impairments induced by PD. DBS consists of surgically connecting one or more electrodes to specific locations in the brain and delivering electrical signals. \cite{okun_deep-brain_2012,nih_dbs} These locations are typically the STN, GPi, and TH due to the substantial role they have in producing our movement. 

One approach that has been taken to advance our comprehension of PD is producing brain network models. These models can provide significant insight into the dynamics of PD, including its progression and the effect of distinct treatments without the risk of performing them on actual patients. \cite{eliasmith_use_2014} Moreover, models of the brain network enhance our grasp on how various sections communicate and influence each other. This is incredibly important for the improvement of DBS, as the exact mechanism of how it works is not known. \cite{perlmutter_deep_2006} On top of this, uncertainty in which brain areas should be implanted with electrodes for DBS is still disputed. \cite{lozano_deep_2019} Simulating how DBS impacts the brain through computational modelling can give valuable insight and help answer these lingering questions. 

Another current challenge in the field of neuroscience is the construction of comprehensive multiscale models. Although single-scale models have advanced our knowledge of brain-related processes, the results only disclose a simplified perspective. Multiscale brain models allow researchers to gain deeper insight into the complicated mechanisms of the brain and NDDs by amalgamating information from the different levels and simulating how they interact. \cite{sheridan_neuroanatomy_2025,krejcar_multiscale_2025} For example, in the context of PD, seeing how the loss of dopaminergic neurons leads to disruptions in the motor circuit. \cite{krejcar_multiscale_2025,khan2023patient,yan2024unraveling}
Accounting for the inherent randomness that is involved in our neuronal activity (i.e. noise) can also clarify how the brain operates and shed light on the progression of PD. \cite{shaheen_neural_2024} This random behaviour can affect processes, such as action potential timing and membrane voltage fluctuations, influencing the overall neural system. \cite{scarciglia_physiological_2025} Including stochastic terms within the model provides a more accurate representation of the brain and potentially more realistic results. Lastly, the movement of ions across brain regions plays an integral role in the communication between neurons and is fundamental for neural functioning. \cite{sykova_diffusion_2008} This study attempts to address these obstacles by introducing a novel model of the BGTH system with large- and small-scale representations of diffusion and stochasticity. In addition, it considers a subnetwork of neurons to mimic the presynaptic inputs that cells within the brain receive. Through this model, we aim to more clearly express the dynamics of the brain in both the healthy and Parkinsonian condition and elucidate the effects of deep brain stimulation. 


\section{\label{sec:method}Model Design of the Brain Network}

\subsection{\label{method:modified_HH}Modified Hodgkin-Huxley Model}

Before introducing the discrete brain network model that is used to depict the membrane potentials of the four regions of focus of the BGTH, we first present the modified Hodgkin-Huxley (HH) equations that was proposed by Maama et al. \cite{hodgkin_quantitative_1952,maama_emergent_2024} The original HH model was developed to describe the propagation of action potentials in a giant squid axon. Since then, the HH model has served as the backbone of multiple mathematical frameworks. \cite{fitzhugh_impulses_1961,hindmarsh_model_1997} Drawing from recent studies that aimed to express the electrical activity of the primary visual cortex (i.e. the V1), Maama et al. adjusts the HH model by incorporating feed-forward inputs through a stochastic input drive and recurrent circuitry by considering excitatory and inhibitory inputs from presynaptic neurons. \cite{maama_emergent_2024,chariker_emergent_2015,chariker_orientation_2016,chariker_rhythm_2018,rangan_emergent_2013,zhou_spatiotemporal_2013} The following ordinary differential equations (ODEs) denote the electrical activity of a single neuron $i$:

\begin{subequations} \label{eq:1}
\begin{align}
\begin{split}
C\frac{dV_{i}}{dt} ={}& \bar{g}_{Na}m^3h(E_{Na} - V_i) + \bar{g}_Kn^4(E_K-V_i)\\
&+\bar{g}_L(E_L-V_i) +g_E(E_E-V_i)+g_I(E_I-V_i)
\label{subeq:1a}
\end{split}\\
\begin{split}
\frac{dn_{i}}{dt} ={}& \alpha_n(V_i)(1-n_i)-\beta_n(V_i)n_i 
\label{subeq:1b}
\end{split}\\
\begin{split}
\frac{dm_{i}}{dt} ={}& \alpha_m(V_i)(1-m_i)-\beta_m(V_i)m_i 
\label{subeq:1c}
\end{split}\\
\begin{split}
\frac{dh_{i}}{dt} ={}& \alpha_h(V_i)(1-h_i)-\beta_h(V_i)h_i 
\label{subeq:1d}
\end{split}\\
\begin{split}
\tau_E\frac{dg_{Ei}}{dt} ={}& -g_{Ei} + S^{dr} \sum\nolimits_{s \in D(i)} \delta(t-s) \\
&+ S^{QE}\sum\nolimits_{j \in\Gamma_E(i), s\in N(j)} \delta(t-s) 
\label{subeq:1e}
\end{split}\\
\begin{split}
\tau_I\frac{dg_{Ii}}{dt} ={}& -g_{Ii} + S^{QI} \sum\nolimits_{j \in \Gamma_I(i)_, s \in N(j)} \delta(t-s)
\label{subeq:1f}
\end{split}
\end{align}
\end{subequations}
with $C=1, \: E_K = -77, \: E_{Na} = 50, \: E_L = -54.387, \: \bar{g}_K = 36, \: \bar{g}_{Na} = 120, \: \text{and} \: \bar{g}_L = 0.3$. In Eq. \ref{subeq:1a}, $V$ is the membrane potential, $g_E$ is the maximum excitatory conductance, $g_I$ represents the maximum inhibitory conductance, and $E_E$ and $E_I$ are the Nernst equilibrium potentials for the excitatory and inhibitory currents. $C$ refers to the membrane capacitance. Eqs. \ref{subeq:1b}\textbf{-}\ref{subeq:1d} are for the gating variables that model the opening and closing of ionic channels. \cite{maama_emergent_2024} More specifically, $n$ depicts the movement of potassium ions across the membrane and $m$ and $h$ represent the sodium fluxes. 

In Eq. \ref{subeq:1e}, the excitatory conductance is based on two factors. Firstly, there are random poissonian kicks of size $S^{dr}$, with $s \in D(i)$ corresponding to the times that $i$ receives these kicks and the Dirac term $\delta(t-s)$ illustrating the instantaneous jumps in conductance that these kicks provide. Secondly, inputs coming from neurons that are presynaptic to $i$ also affect $g_{Ei}$. The sizes of these kicks are of magnitude $S^{EE}$ or $S^{IE}$ depending on whether the neuron $i$ is excitatory or inhibitory. The number of presynaptic neurons that contribute these inputs are described by $j \in\Gamma_E(i) \text{ and } s\in N(j)$. $j \in\Gamma_E(i)$ portrays the set of all excitatory neurons to $i$ and $s\in N(j)$ are the neurons that have crossed the threshold to spike and consequently add to the excitatory conductance. Additionally, $\tau_E$ depicts the decay time constant for $g_{Ei}$. The presynaptic excitatory current is expressed as $I_{Ei} = g_{Ei}(E_E - V_i)$. Similarly, the inhibitory conductance in Eq. \ref{subeq:1f} includes kicks of size $S^{EI}$ and $S^{II}$ coming from the connected inhibitory neurons contingent on if $i$ is excitatory or inhibitory. The summation term is identical to that of $g_{Ei}$ except that the set of all inhibitory neurons to $i$ are considered, not excitatory. There is also no stochastic input drive for the inhibitory conductance. Additionally, $\tau_I$ is the decay time constant for $g_{Ii}$. The presynaptic inhibitory current is modelled as $I_{Ii} = g_{Ii}(E_I - V_i)$.

\subsection{\label{method:discrete_model}Discrete Brain Network: Spatio-Temporal Domain}
\begin{figure}
\includegraphics[scale=0.35]{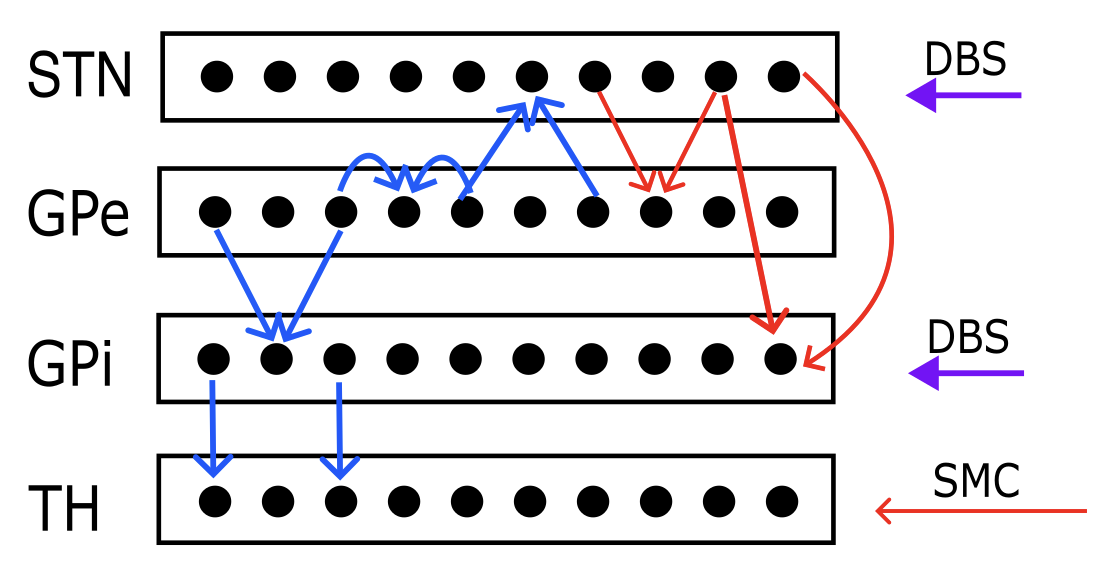}
\caption{\label{fig:nuclei}(Color online) The BGTH network model. The red arrows represent excitatory the connections and the blue arrows denote the inhibitory inputs. There is excitatory input from the SMC being projected to the TH. The purple arrows portray the DBS stimulus to the STN and GPi.}
\end{figure}

We offer a novel approach of modelling the BGTH network by synthesizing the method used in [\onlinecite{shaheen_multiscale_2022}] and Eqs. \ref{subeq:1a}\textbf{--}\ref{subeq:1f}.\cite{ maama_emergent_2024} The model alters the framework introduced by Rubin and Terman, and comprises the four main nuclei in the BGTH: the STN, GPi, GPe, and the TH.\cite{rubin2004high,lu2019application} These four regions are linked by several excitatory and inhibitory connections which the model considers, and responds to information stemming from the sensorimotor cortex (SMC). Note that these associations are derived from known physiological connections which can be observed in Fig. \ref{fig:nuclei}. We harnessed the same reaction kinetics and equations for the gating variables applied in [\onlinecite{shaheen_multiscale_2022}] which are described by Eqs. \ref{gating:1}\textbf{--}\ref{gating:5}. The method outlined here follows the graph-network approach in [\onlinecite{shaheen_multiscale_2022}] which takes inspiration from Thompson et al. to capture the diffusion dynamics across the brain network. \cite{thompson2020protein} The connectome is defined by a graph $G$ with $V$ nodes and $E$ edges based on diffusion tensor imaging (DTI) and tractography data from the Human Connectome Project (HCP) to encapsulate the change of voltage over time at different nodal points.\cite{thompson2020protein,mcnab_human_2013,szalkai2017parameterizable,kerepesi2016direct} In greater detail, $W$ is the weighted adjacency matrix that mirrors Eq. \ref{W}. The term $n_{ij}$ is the mean fiber number and $l^2_{ij}$ is the mean length squared of the axonal bundles that connect the nodes $i$ and $j$. The weighted degree matrix $D$, follows Eq. \ref{D}. These terms culminate in the weighted graph Laplacian, $L$ in Eq. \ref{laplacian}, where $\rho$ is the diffusion coefficient. This Laplacian operator shapes the diffusion dynamics of the system.

\begin{figure}
    \includegraphics[scale=0.22]{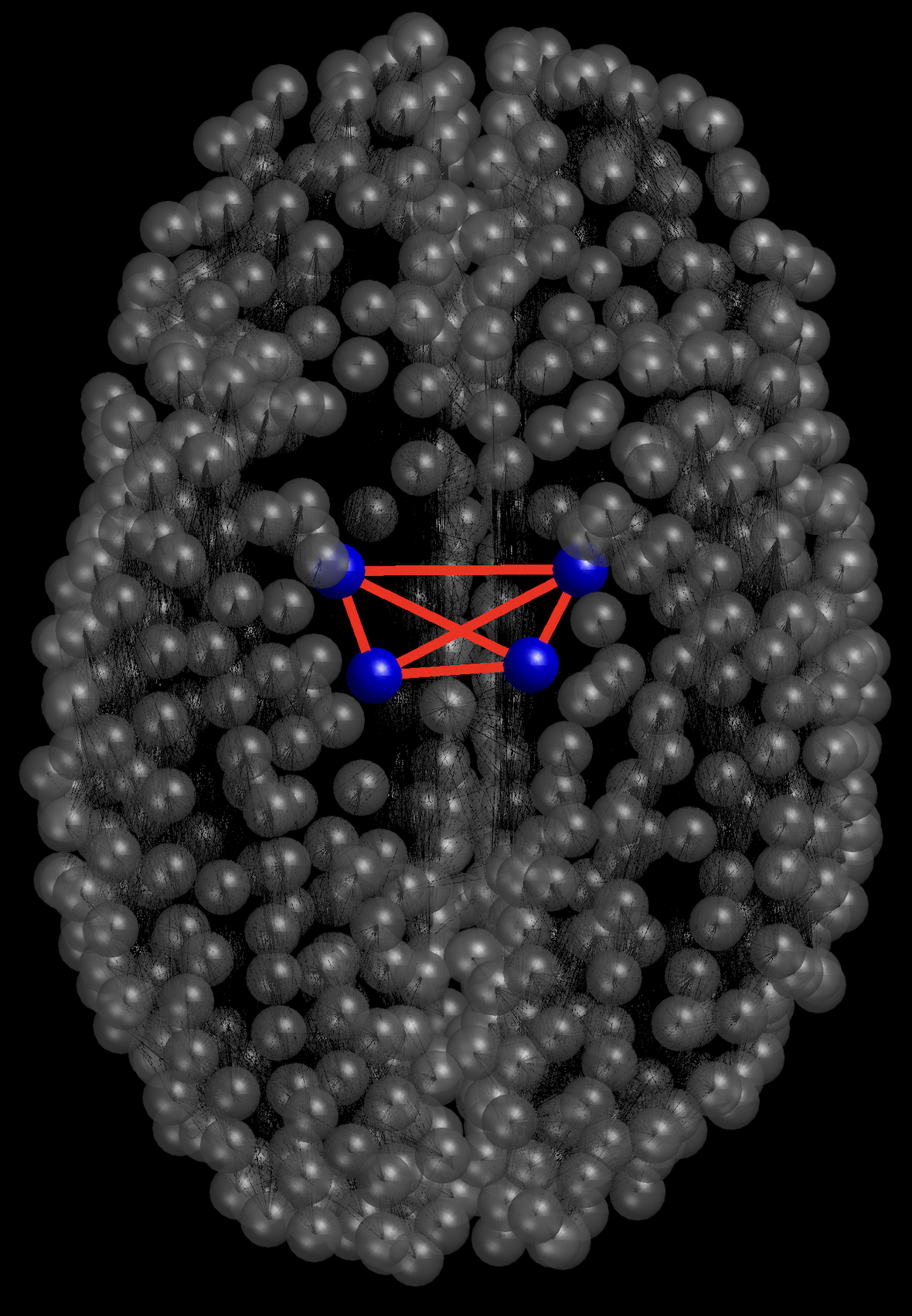}
    \caption{\label{fig:nodalnetwork}(Color online) The brain network connectome in the healthy state. The blue nodes represent the STN, GPe, GPi, and TH. The red connections between these nodes indicate the diffusion between these regions.}
\end{figure}

\begin{subequations} \label{gating_variables}
\begin{align}
\frac{d{h^p}}{d{t}} &= e_{h^p}[h^p_\infty - h^p]/\tau_{h^p}&& \label{gating:1}\\
\frac{d{n^p}}{d{t}} &= e_{n^p}[n^p_\infty - n^p]/\tau_{n^p}&& \label{gating:2}\\
\frac{d{r^p}}{d{t}} &= e_{r^p}[r^p_\infty - r^p]/\tau_{r^p}&& \label{gating:3}\\
\frac{d{c^p}}{d{t}} &= e_{c^p}[c^p_\infty - c^p]/\tau_{c^p}&& \label{gating:4}\\
\frac{d{w^p}}{d{t}} &= e_{w^p}(-I_{C_a} - I_T - l_{w^p}w^p)&& \label{gating:5}
\end{align}
\end{subequations}

\begin{align}
W      & = \frac{n_{ij}}{l^2_{ij}}, \quad i,j = 1,\dots,V && \label{W} \\
D_{ii} & = \sum\nolimits^V_{j=1} W_{ij}, \quad i = 1,\dots,V && \label{D} \\
L_{ij} & = \rho(D_{ij} - W_{ij}), \quad i,j = 1,\dots,V && \label{laplacian}
\end{align}

In order to merge the modified HH model established in [\onlinecite{maama_emergent_2024}] with this discrete BGTH network,  a system of fifty neurons following Eqs. \ref{subeq:1a}\textbf{--}\ref{subeq:1f} for the STN, GPe, GPi, and TH were simulated. The number of excitatory and inhibitory neurons for each region was determined by the ratio used in [\onlinecite{maama_emergent_2024}]. At each time step for $dt = 0.01$ ms, the average excitatory and inhibitory currents from the neuronal network were added to the equations for the membrane voltage of the corresponding brain regions. The equations for the spatio-temporal model are:

\begin{subequations} \label{discrete_network_eq}
\begin{align}
\begin{split}
\frac{d{v_{1}}}{d{t}} ={}& -d_{v^{sn}} \sum^V_{k=1}L_{1k}v_k + \frac{1}{c_m}(-I^{sn}_{Na} - I^{sn}_K-I^{sn}_L-I^{sn}_T
-I^{sn}_{Ca}\\
& -I^{sn}_{ahp}-I_{ge \rightarrow sn} + I_{snapp} + \bar{I}_{E,sn} + \bar{I}_{I,{sn}}) 
\label{subeq:2a}
\end{split}\\
\begin{split}
\frac{d{v_{2}}}{d{t}} ={}& -d_{v^{gi}} \sum^V_{k=1}L_{2k}v_k + \frac{1}{c_m}(-I^{gi}_{Na} - I^{gi}_K-I^{gi}_L-I^{gi}_T
-I^{gi}_{Ca}\\
& -I^{gi}_{ahp}-I_{sn \rightarrow gi} - I_{ge \rightarrow gi}+ I_{giapp} + \bar{I}_{E,{gi}} + \bar{I}_{I,{gi}}) 
\label{subeq:2b}
\end{split}\\
\begin{split}
\frac{d{v_{3}}}{d{t}} ={}& -d_{v^{ge}} \sum^V_{k=1}L_{3k}v_k + \frac{1}{c_m}(-I^{ge}_{Na} - I^{ge}_K-I^{ge}_L-I^{ge}_T
-I^{ge}_{Ca}\\
& -I^{ge}_{ahp}-I_{sn \rightarrow ge} - I_{ge \rightarrow ge} + I_{geapp} + \bar{I}_{E,{ge}} + \bar{I}_{I,{ge}}) 
\label{subeq:2c}
\end{split}\\
\begin{split}
\frac{d{v_{4}}}{d{t}} ={}& -d_{v^{th}} \sum^V_{k=1}L_{4k}v_k + \frac{1}{c_m}(-I^{th}_{Na} - I^{th}_K-I^{th}_L-I^{th}_T-I_{gi \rightarrow th}\\
& + I_{smc} + \bar{I}_{E,{th}} + \bar{I}_{I,{th}}) 
\label{subeq:2d}
\end{split}
\end{align}
\end{subequations}

\begin{subequations} \label{avg_current}
\begin{align}
\bar{I}_{E,p} &= \frac{1}{N_p}\sum_{i=1}^{N_p}g_{Ei,{p}}(E_E-V_{i,p})\label{IEp}, \quad p = sn, gi, ge, th&&\\
\bar{I}_{I,p} &= \frac{1}{N_p}\sum_{i=1}^{N_p}g_{Ii,{p}}(E_I-V_{i,p})\label{IIp}, \quad p = sn, gi, ge, th&&
\end{align}
\end{subequations}
where $\bar{I}_{E,{p}} \text{ and } \bar{I}_{I,{p}}$ are the average excitatory and inhibitory currents for the neurons in region $p$, $p = sn, gi, ge, th$ and follow Eqs. \ref{IEp}\textbf{-}\ref{IIp}. The terms $v_1$, $v_2$, $v_3$, and $v_4$ are the membrane potentials for the STN, GPi, GPe, and TH neurons. Also, $d_{v^p}$ is the diffusion term for the node for region $p$, with $N_p$ being the number of neurons for that particular region. As previously mentioned, $N_p = 50$ for all $p$. To solve this model, Euler’s method was used. In addition, the PD state was simulated by setting the parameter $pd = 1$ and $pd = 0$ when in the healthy state which aligns with the work from Lu et al.\cite{lu2019application} This reduces the constant bias currents $I_{app}$ to the STN, GPi, and GPe neurons. Therefore, we have constructed an original co-simulation model that focuses on the cortex-BGTH network. It encompasses the sophisticated diffusion dynamics across the entire brain and considers the recurrent input of presynaptic neurons and a feed-forward stochastic drive.

\subsection{Adding Deep Brain Stimulation}
To simulate the effects of DBS on the PD state, a DBS current term denoted $I_{DBS}$ is added to Eqs. \ref{subeq:2a}\textbf{--}\ref{subeq:2d}. Similar to the work by Shaheen et al., this term is only added to the STN and GPi equations as they are the primary targets for DBS application. \cite{shaheen_multiscale_2022} The DBS term is specifically modelled as $I_{DBS} = i_DH(sin(2\pi t/\rho_D))\cdot[1-H(sin(2\pi(t+\delta_D)/\rho_d))]$, where $i_D = 200 \mu A/\text{cm}^2$ is the amplitude of the DBS stimulation, $\delta_D = 0.6$ ms is the impulse length, and $\rho_D = 6$ is the stimulation period which is integrated from [\onlinecite{lu2019application,shaheen_deep_2022}]. It is important to mention that this is an open-loop application. The equations for the STN and GPi membrane potentials in the brain network model with the DBS term are:

\begin{subequations} \label{DBS_network_eqs}
\begin{align}
\begin{split}
\frac{d{v_{1}}}{d{t}} ={}& -d_{v^{sn}} \sum^V_{k=1}L_{1k}v_k + \frac{1}{c_m}(-I^{sn}_{Na} - I^{sn}_K-I^{sn}_L-I^{sn}_T-I^{sn}_{Ca}\\
&-I^{sn}_{ahp}-I_{ge \rightarrow sn} + I_{snapp} + I_{DBS} + \bar{I}_{E,{sn}} + \bar{I}_{I,{sn}}) 
\label{eq:DBS_STN}
\end{split}\\
\begin{split}
\frac{d{v_{2}}}{d{t}} ={}& -d_{v^{gi}} \sum^V_{k=1}L_{2k}v_k + \frac{1}{c_m}(-I^{gi}_{Na} - I^{gi}_K-I^{gi}_L-I^{gi}_T-I^{gi}_{Ca}-I^{gi}_{ahp}\nonumber\\
&-I_{sn \rightarrow gi} - I_{ge \rightarrow gi}+ I_{giapp} + I_{DBS} + \bar{I}_{E,{gi}} + \bar{I}_{I,{gi}}) 
\label{eq:DBS_GPi}
\end{split}\\
\end{align}
\end{subequations}

\subsection{Implementing Stochastic Noise}
Adopting the approach from Shaheen and Melnik, stochastic noise is added to the discrete brain network.\cite{shaheen_neural_2024} The new equations are:
\begin{subequations} \label{stochastic_network_eqs}
\begin{align}
\begin{split}
\frac{d{v_{1}}}{d{t}} ={}& -d_{v^{sn}} \sum\nolimits^V_{k=1}L_{1k}v_k + \frac{1}{c_m}(-I^{sn}_{Na} - I^{sn}_K-I^{sn}_L-I^{sn}_T-I^{sn}_{Ca}\\
&-I^{sn}_{ahp}-I_{ge \rightarrow sn} + I_{snapp} + \bar{I}_{E,{sn}} + \bar{I}_{I,{sn}}) \\
&+ \sigma_1 \cdot dW_1(t)
\end{split}\\
\begin{split}
\frac{d{v_{2}}}{d{t}} ={}& -d_{v^{gi}} \sum\nolimits^V_{k=1}L_{2k}v_k + \frac{1}{c_m}(-I^{gi}_{Na} - I^{gi}_K-I^{gi}_L-I^{gi}_T-I^{gi}_{Ca}\\
&-I^{gi}_{ahp}-I_{sn \rightarrow gi} - I_{ge \rightarrow gi}+ I_{giapp} + \bar{I}_{E,{gi}} + \bar{I}_{I,{gi}}) \\
&+ \sigma_2 \cdot dW_2(t)
\end{split}\\
\begin{split}
\frac{d{v_{3}}}{d{t}} ={}& -d_{v^{ge}} \sum\nolimits^V_{k=1}L_{3k}v_k + \frac{1}{c_m}(-I^{ge}_{Na} - I^{ge}_K-I^{ge}_L-I^{ge}_T-I^{ge}_{Ca}\\
&-I^{ge}_{ahp}-I_{sn \rightarrow ge} - I_{ge \rightarrow ge} + I_{geapp} + \bar{I}_{E,{ge}} + \bar{I}_{I,{ge}})\\
&+ \sigma_3 \cdot dW_3(t)
\end{split}\\
\begin{split}
\frac{d{v_{4}}}{d{t}} ={}& -d_{v^{th}} \sum\nolimits^V_{k=1}L_{4k}v_k + \frac{1}{c_m}(-I^{th}_{Na} - I^{th}_K-I^{th}_L-I^{th}_T-I_{gi \rightarrow th} \\
&+ I_{smc} + \bar{I}_{E,{th}} + \bar{I}_{I,{th}}) + \sigma_4 \cdot dW_4(t)
\end{split}
\end{align}
\end{subequations}
where $\sigma_i, \, i = 1,2,3,4$ are the scaling factors which control the noise intensity for the four brain regions of interest. Their values are $\sigma_1 = 0.1$, $\sigma_2 = 0.4$, $\sigma_3 = 0.4$, and $\sigma_4 = 0.5$. Furthermore, $dW_i(t)$ is the increment of the Wiener process $W_i(t)$ where $dt=0.01$ ms is the size of the increment. Due to the added white noise, these stochastic differential equations (SDEs) are solved using the Euler-Maruyama method. It is worth mentioning that when DBS is employed, the only difference is that $I_{DBS}$ is added to the STN and GPi equations similar to Eqs. \ref{eq:DBS_STN}\textbf{--}\ref{eq:DBS_GPi}.


\section{\label{results}Results}
\subsection{\label{results:healthy&pd}The Healthy and Parkinsonian State}
\begin{figure}
\includegraphics[scale=0.63]{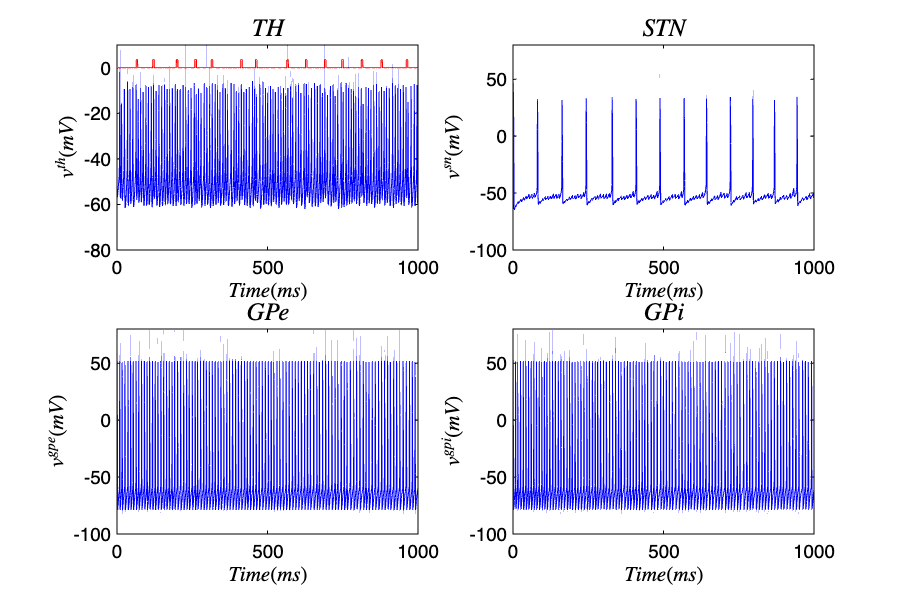}
\caption{\label{healthy_spiking}(Color online) Membrane voltages of the TH, STN, GPe, and GPi neurons in a healthy state for the discrete
brain network. The red pulse train refers to SMC signals.}
\end{figure}

\begin{figure}
\includegraphics[scale=0.63]{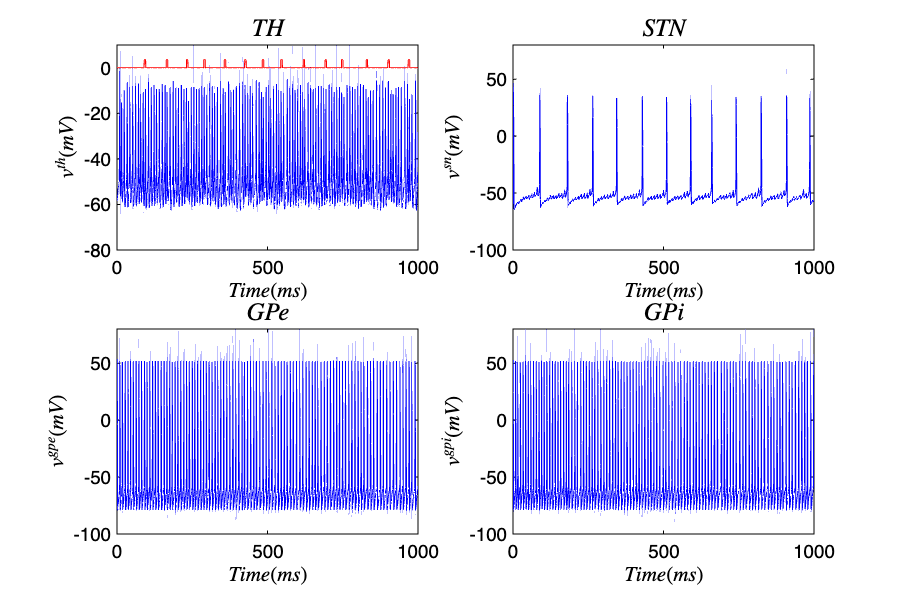}
\caption{\label{healthy_spiking_stochastic}(Color online) Membrane voltages of the TH, STN, GPe, and GPi neurons in a healthy state for the discrete
brain network with stochastic noise. The red pulse train refers to SMC signals.}
\end{figure}

\begin{figure}
\includegraphics[scale=0.63]{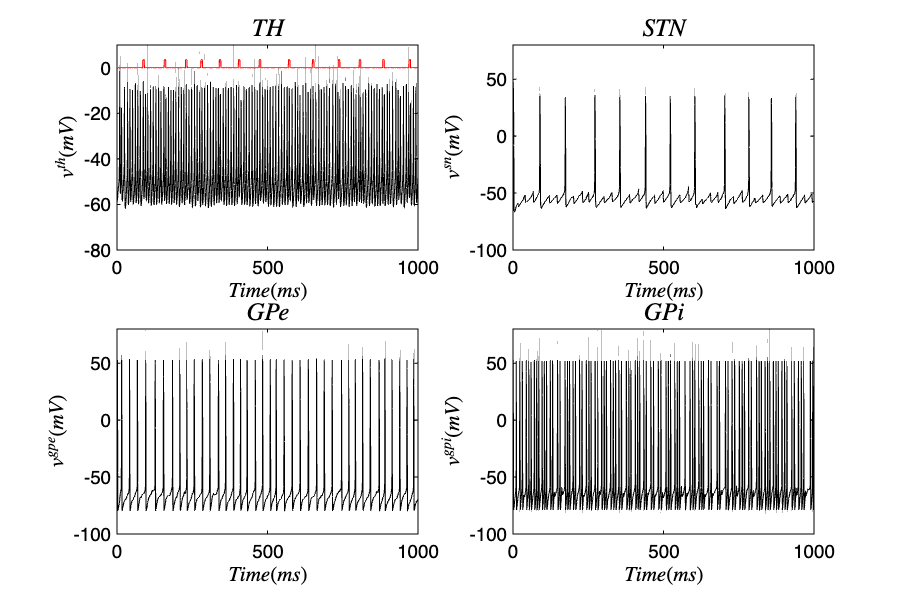}
\caption{\label{PD_spiking}(Color online) Membrane voltages of the TH, STN, GPe, and GPi neurons in the Parkinsonian state for the discrete brain network. The red pulse train refers to SMC signals.}
\end{figure}

\begin{figure}
\includegraphics[scale=0.63]{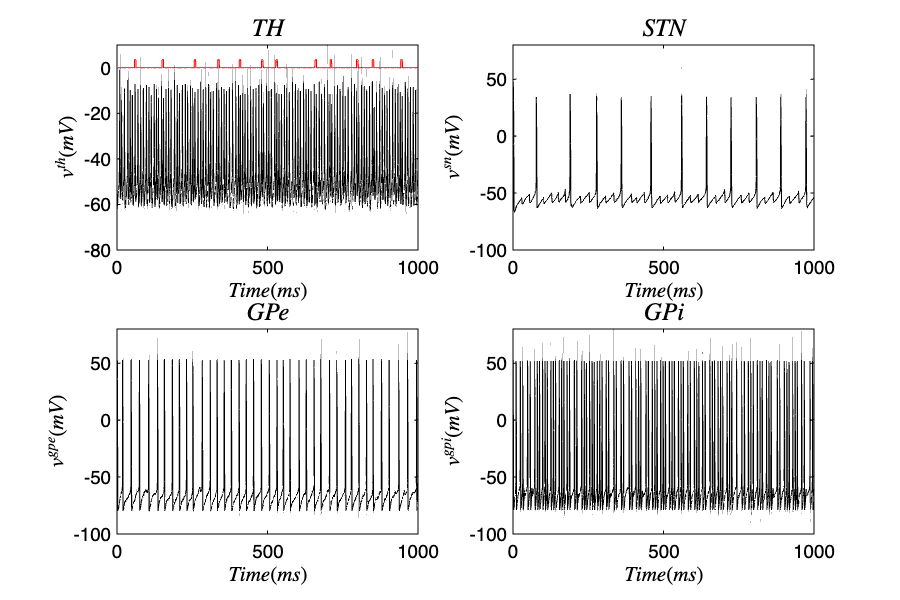}
\caption{\label{PD_spiking_stochastic}(Color online) Membrane voltages of the TH, STN, GPe, and GPi neurons in the Parkinsonian state for the discrete brain network with stochastic noise. The red pulse train refers to SMC signals.}
\end{figure}

\begin{figure}
\includegraphics[scale=0.6]{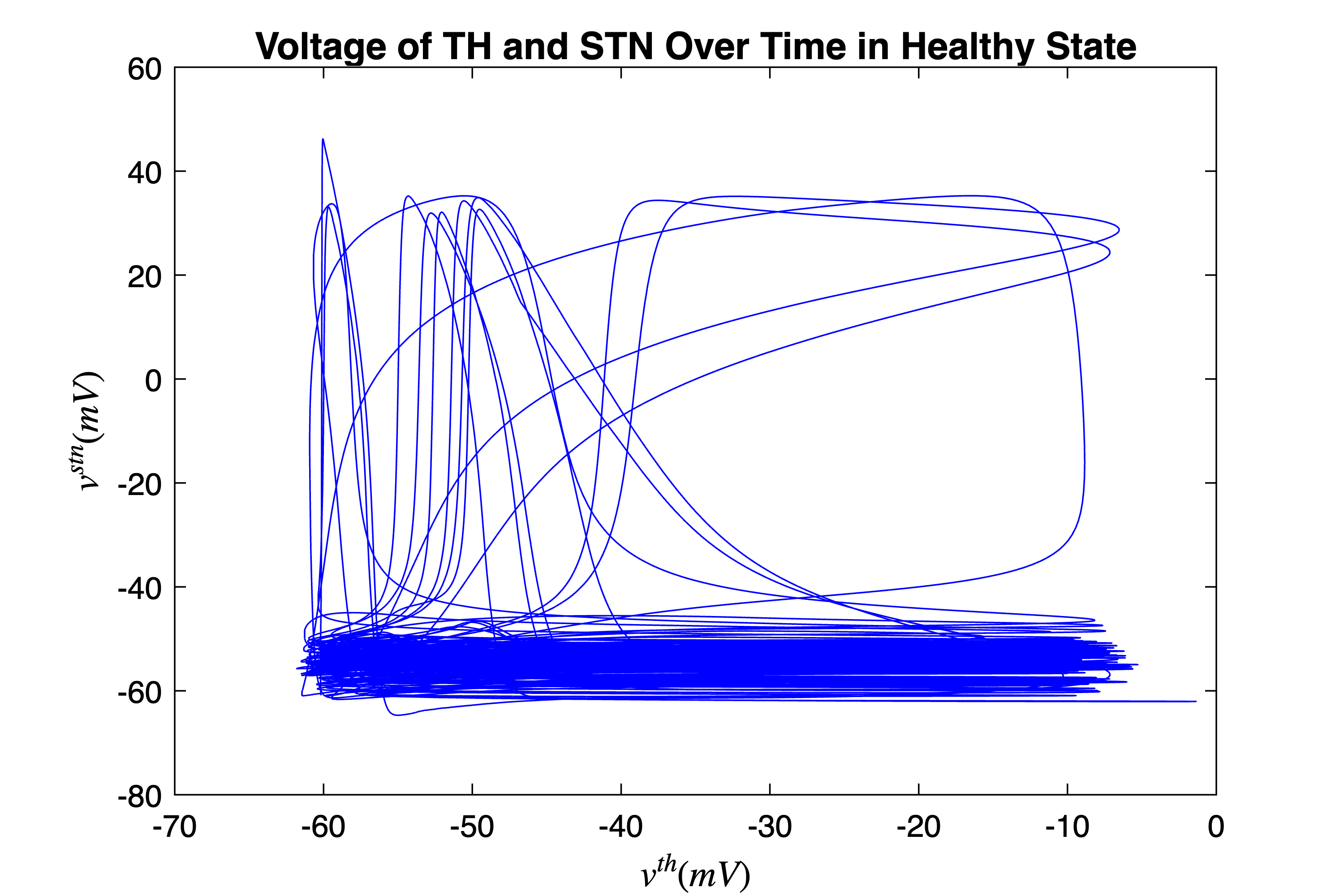}
\caption{\label{healthy_nontemp}(Color online) Membrane voltages of the TH and STN neurons in a healthy state for 1000 ms.}
\end{figure}

\begin{figure}
\includegraphics[scale=0.6]{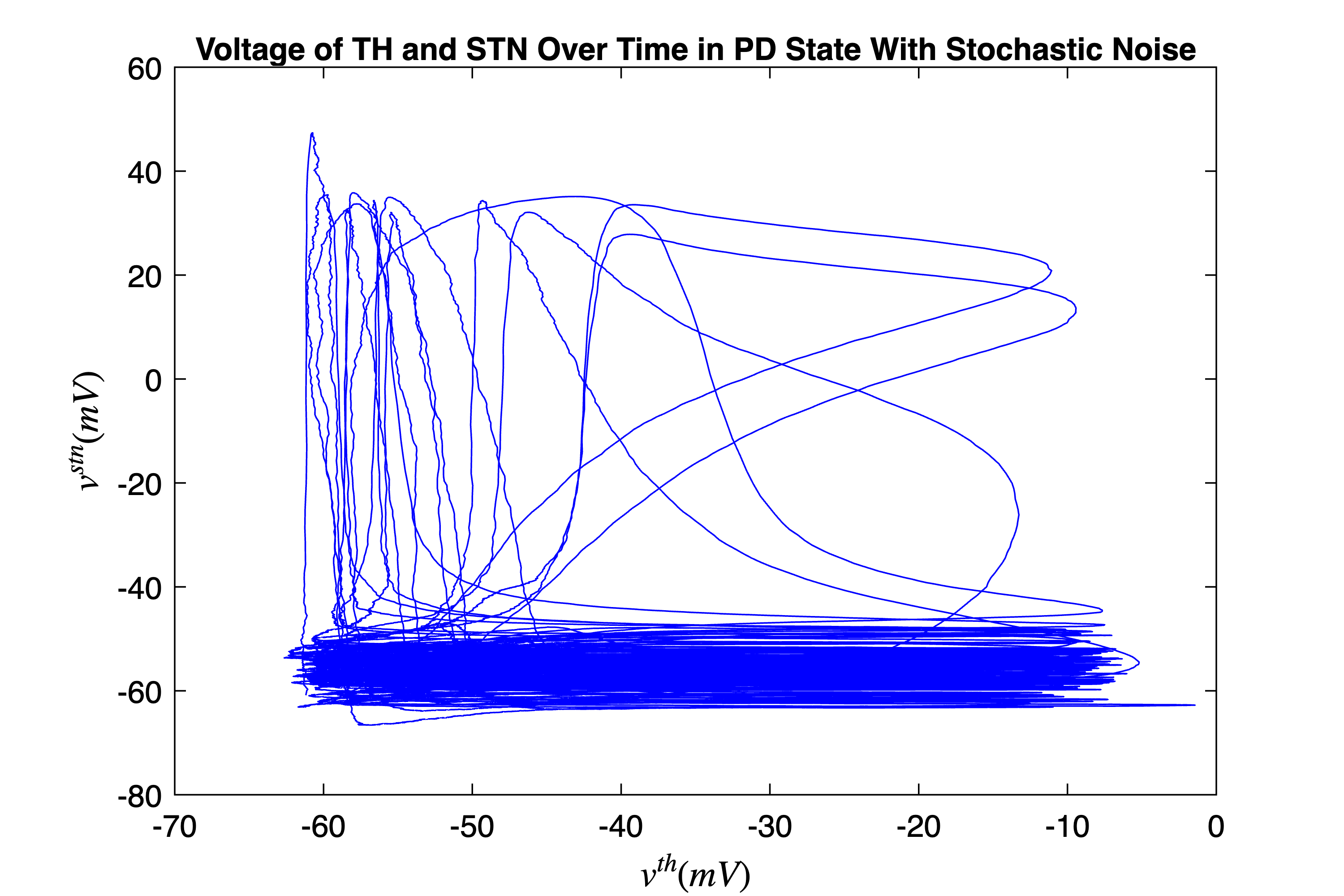}
\caption{\label{PD_nontemp_stochastic}(Color online) Membrane voltages of the TH and STN neurons in the Parkinsonian state for 1000 ms. No DBS was applied and stochastic noise was present.}
\end{figure}

Fig. \ref{healthy_spiking} and Fig. \ref{healthy_spiking_stochastic} portray the healthy state of the brain when stochastic noise was absent and present. Fig. \ref{PD_spiking} and Fig. \ref{PD_spiking_stochastic} reflect the membrane voltages of the TH, STN, GPe, and GPi while in the Parkinsonian state. By comparing the spiking activity, we found that healthy neurons exhibit greater oscillatory patterns between spikes than neurons in the PD state. We also found that the healthy GPe neurons displayed more spikes than when diseased. These findings were consistent in the deterministic and stochastic model. Additionally, certain instances of the simulation revealed exaggerated spike jumps when compared to the typical spiking activity. For example, in Fig. \ref{healthy_spiking} the average spikes for the TH region hovered around -10 mV. However, several spikes leaned closer to 0 mV and some even spiking to positive values. These irregular, greater spikes also occurred in the other regions. Interestingly, similar behaviour in the TH was still present in the new model even without stochastic noise being added. Moreover, from the aforementioned figures, it is clear that the thalamic spikes were not uniform and achieved varying levels of voltage. It is worth noting that the ability of the TH to accurately respond to information from the SMC was disrupted in both cases, as evidenced by the TH spiking without explicit SMC input. Figures \ref{healthy_nontemp}\textbf{--}\ref{PD_nontemp_stochastic} illustrate how the voltage patterns of the TH and STN changed during the simulation. In the healthy and PD condition, the voltages across both regions displayed synchronized and unsynchronized spiking. Additionally, the added stochastic noise resulted in a more “fuzzy” pattern in comparison to the smoother lines in the figures that denote the deterministic approach. Corresponding to the earlier diagrams with high thalamic activity, the TH spiked on many instances where the STN was at or near its resting state. 

\begin{figure}
\includegraphics[scale=0.6]{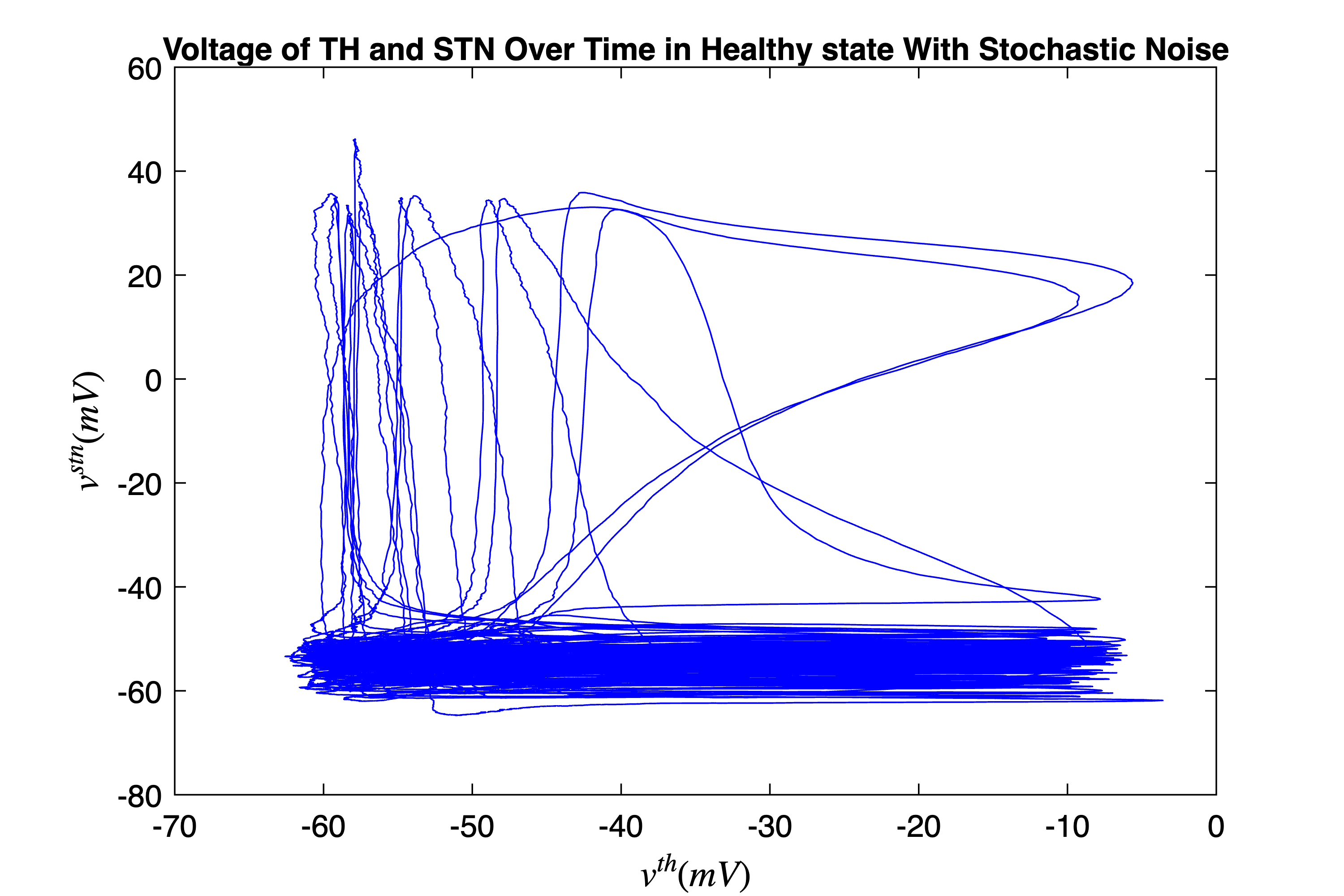}
\caption{\label{healthy_nontemp_stochastic}(Color online) Membrane voltages of the TH and STN neurons in a healthy state for 1000 ms. Stochastic noise was present.}
\end{figure}

\begin{figure}
\includegraphics[scale=0.6]{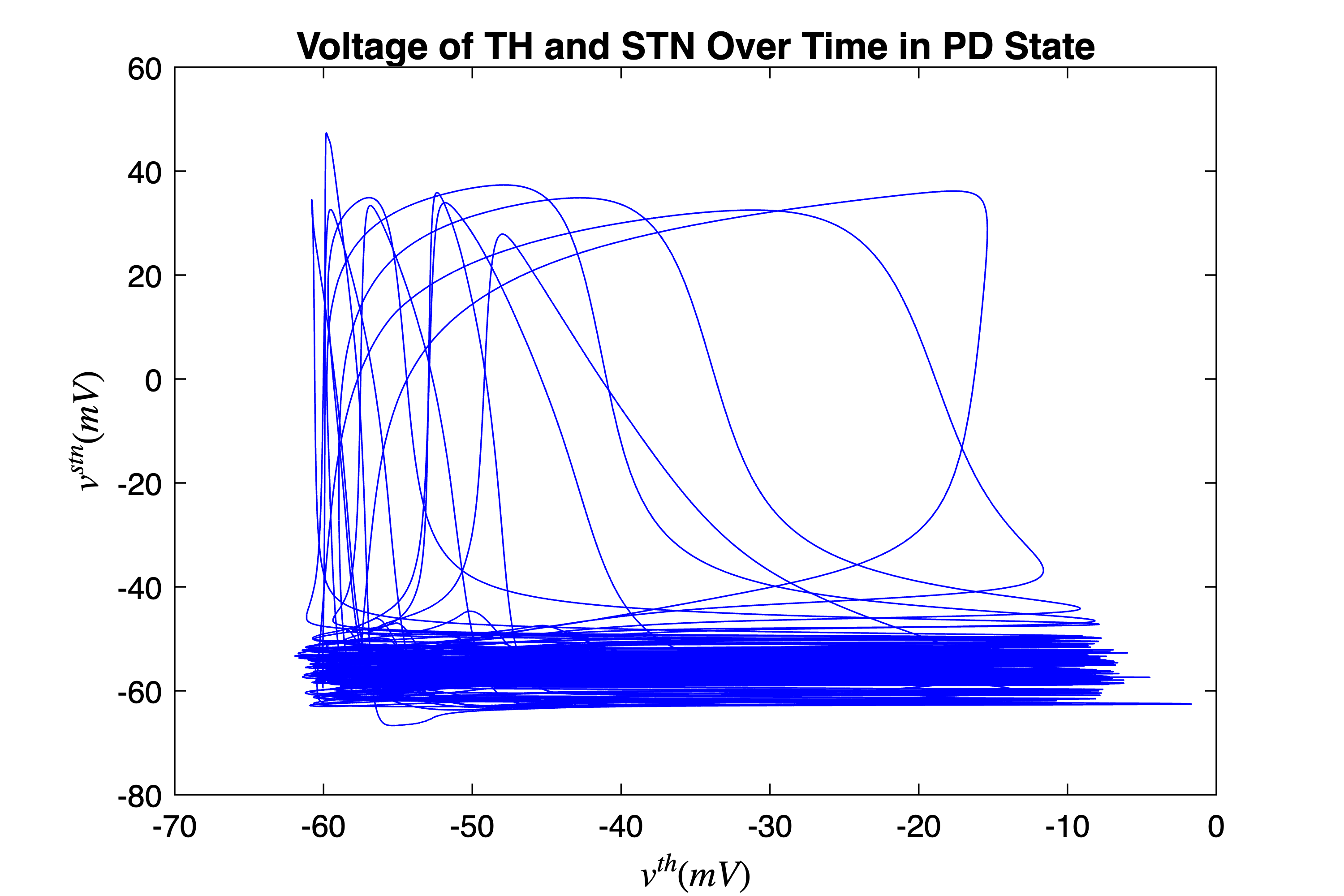}
\caption{\label{PD_nontemp}(Color online) Membrane voltages of the TH and STN neurons in the Parkinsonian state for 1000 ms. No DBS was applied.}
\end{figure}

\subsection{\label{results:dbs}Results of Deep Brain Stimulation}
\begin{figure}
\includegraphics[scale=0.63]{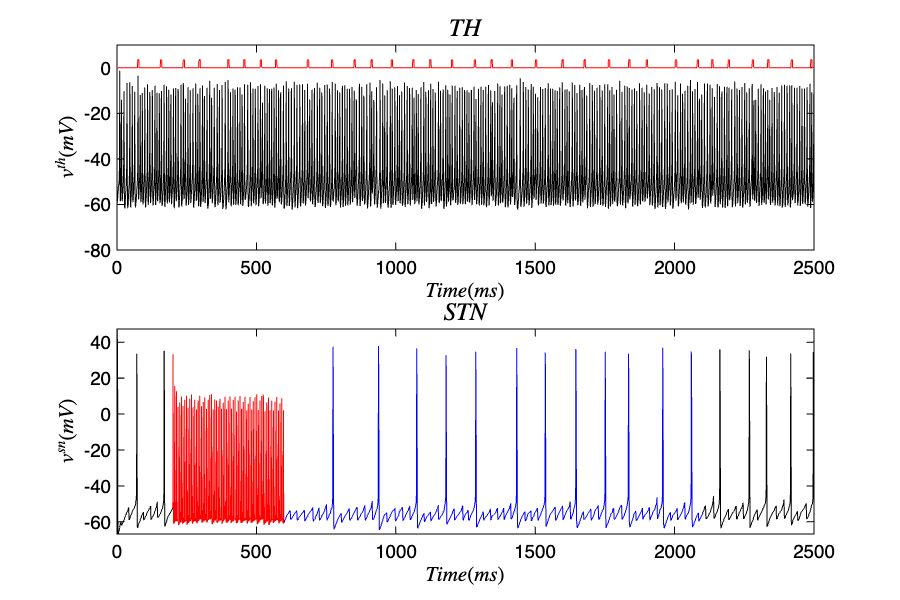}
\caption{\label{dbs_stn_stochastic}(Color online) Membrane voltages of the STN and TH neurons in the Parkinsonian state for the discrete brain network with stochastic noise.
Black represents the PD state, red denotes the application of DBS to the STN, and blue refers to the healthy state.
}
\end{figure}

\begin{figure}
\includegraphics[scale=0.63]{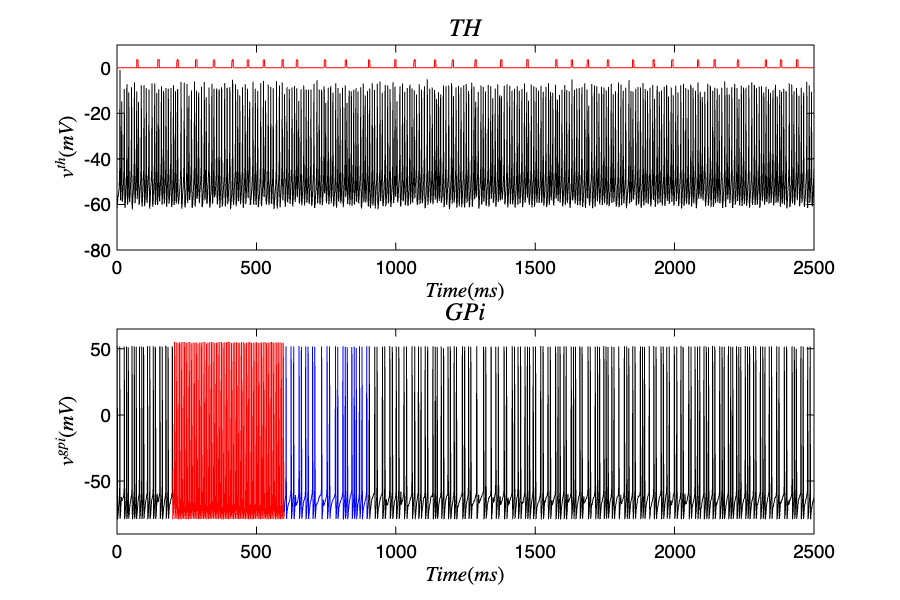}
\caption{\label{dbs_gpi_stochastic}(Color online) Membrane voltages of the GPi and TH neurons in the Parkinsonian state for the discrete brain network with stochastic noise. Black represents
the PD state, red denotes the application of DBS to the GPi, and blue refers to the healthy state.
}
\end{figure}

\begin{figure}
\includegraphics[scale = 0.6]{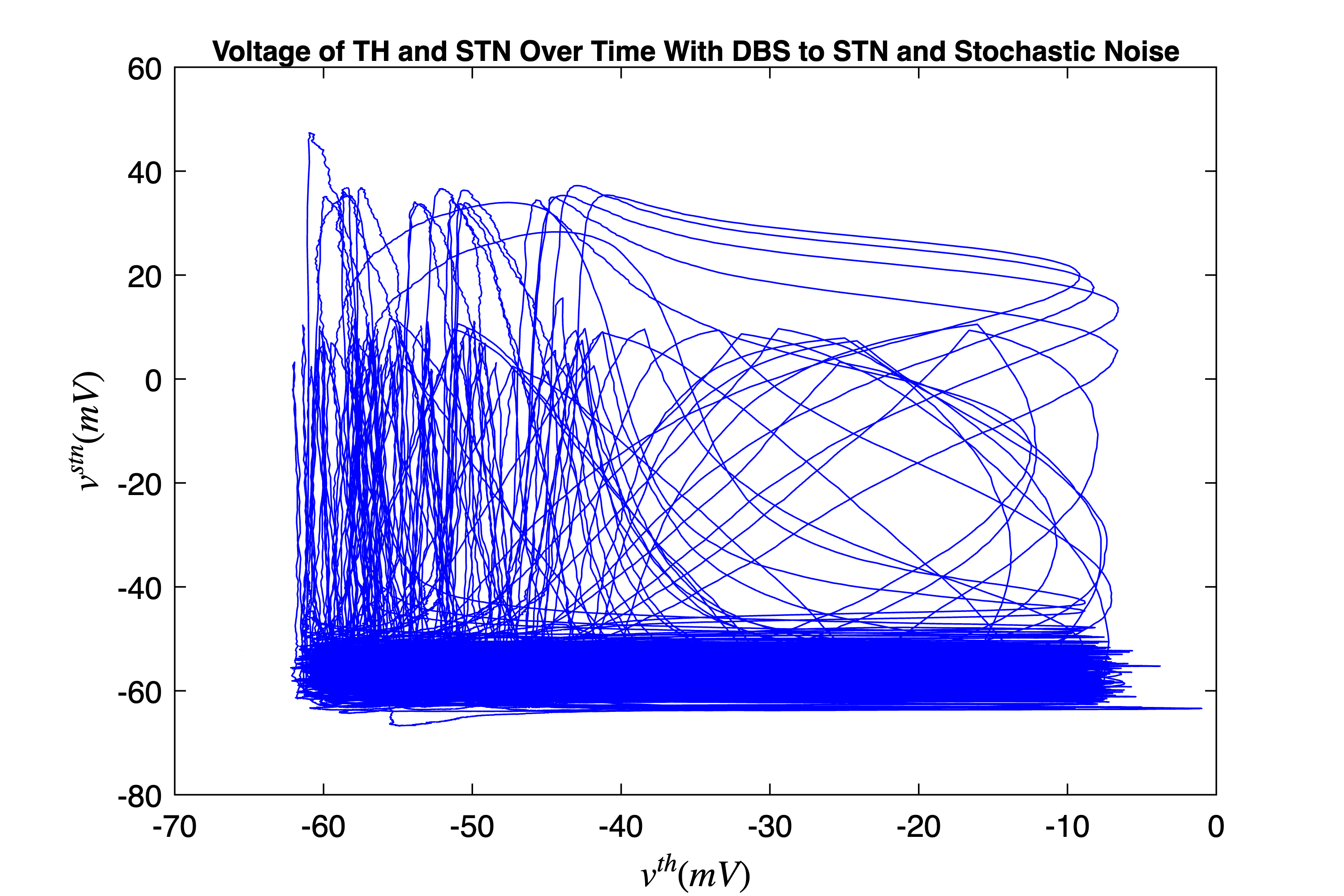}
\caption{\label{nontemp_dbs_stn_stochastic}(Color online) Membrane voltages of the TH and STN neurons in the Parkinsonian state for 2500 ms. DBS was applied to the STN and stochastic noise was present.}
\end{figure}

\begin{figure}
\includegraphics[scale = 0.32]{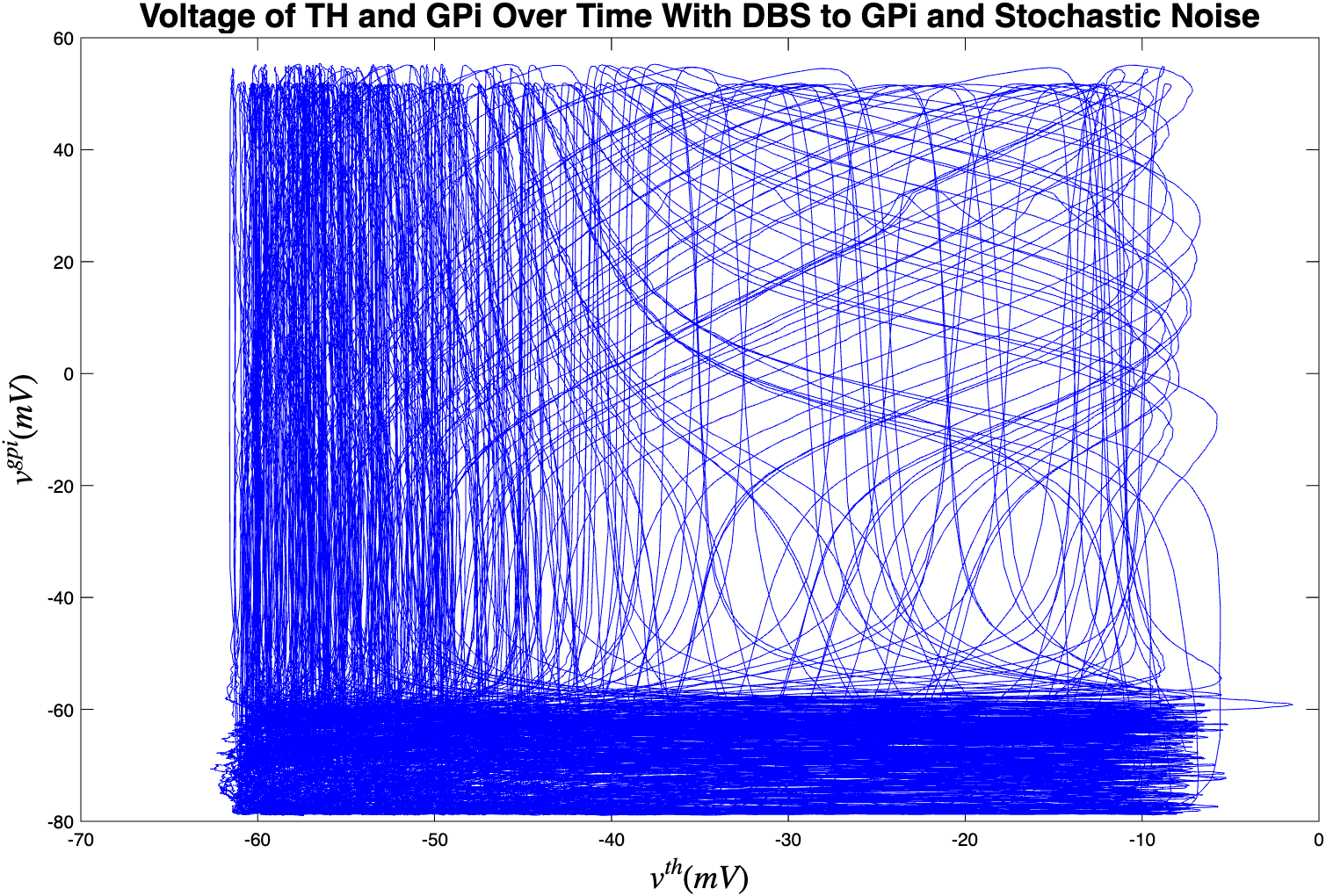}
\caption{\label{nontemp_dbs_gpi_stochastic}(Color online) Membrane voltages of the TH and GPi neurons in the Parkinsonian state for 2500 ms. DBS was applied to the GPi and stochastic noise was present.}
\end{figure}

Next, open-loop DBS was applied to the STN and GPi neurons while in the PD condition which are presented in Fig. \ref{dbs_stn_stochastic} and Fig. \ref{dbs_gpi_stochastic}. After the application of DBS, the STN and GPi neurons were able to transition into the healthy state temporarily. We discovered that the healthy state experienced an increase in the oscillations between spikes and the total spiking activity in the STN and GPi neurons were reduced than when in the diseased condition. Akin to the outcomes in section \ref{results:healthy&pd}, the thalamus had many spikes with unsteady voltage values. The majority of these spikes took place even without information being relayed by the SMC. When DBS was applied to either the STN or GPi, there was more variety in the voltage patterns between the TH and STN, and the TH and GPi as seen in Fig. \ref{nontemp_dbs_stn_stochastic} and Fig. \ref{nontemp_dbs_gpi_stochastic}. The total movement was greater when DBS was applied to the GPi, which coincides with the more extensive spiking in Fig. \ref{dbs_gpi_stochastic}. There were occasions where the spikes happened simultaneously (i.e. when the voltage reaches the top-right corner) or independently while the other region was at rest (i.e. when the voltage is at the top-left corner or bottom-right corner). In both cases where DBS was administered to the STN or GPi, we saw more STN spiking activity which was closely followed by TH spikes compared to the cases where DBS was not applied. This result also materialized with the GPi neurons. Furthermore, many thalamic spikes happened when the STN was at rest and similarly for the GPi. This is demonstrated by the cluster of straight lines at the bottom of Fig. \ref{nontemp_dbs_stn_stochastic} and Fig. \ref{nontemp_dbs_gpi_stochastic}. Another remarkable detail, was that the GPi neurons had a wider range of independent spiking while the TH was rest as opposed to the STN neurons. 

\section{\label{discussion}Discussion}
As previously mentioned in section \ref{results:healthy&pd}, the firing of the TH neurons were characterized by numerous spikes with fluctuating values. This was contrary to the result in [\onlinecite{shaheen_multiscale_2022}] where the TH experienced less, steadier, tonic spiking. Since this inconsistent spiking endured in the deterministic model in both the healthy and PD conditions, we can infer that this behaviour is not caused by noise, but rather, manifests as a result of the high spiking activity in the neuronal subnetwork of the TH (i.e. the neurons following the equations in \ref{method:modified_HH}). Consequently, these neurons generate a stronger presynaptic current that can amplify the membrane voltage of the TH and cause these slightly higher jumps in spiking. A similar trend can also be observed when in the PD state and DBS is applied to the STN or GPi. Another perspective that can explain this result is that since large and irregular thalamic spiking activity emerges in both the deterministic and stochastic models, this suggests that such dynamics are intrinsic to the network connectivity and conductance interactions rather than being solely noise-driven. The inclusion of stochastic terms does not fundamentally alter the overall spiking regime, but instead introduces additional variability in spike timing and amplitude, producing less smooth and more biologically realistic trajectories. This effect is most clearly observed as a “fuzzier” pattern in the stochastic simulations compared to the smoother deterministic outputs, highlighting noise as a modulator of existing variability rather than a primary cause of spiking. In contrast to prior work, where noise was reported to be essential for disrupting thalamic activity, our findings demonstrate that irregular dynamics persist even in its absence, emphasizing the robustness of the model and suggesting that stochasticity serves mainly to enhance variability and better capture the inherent unpredictability of neural firing. \cite{shaheen_neural_2024} With these differing outcomes on the role of stochastic noise, further research should continue to explore whether stochasticity acts as a catalyst for greater brain activity and potentially brain dysfunction, or mainly a component that fosters the flexibility in neural dynamics. 

Although the framework presented in this study provides a novel approach for modelling the cortex-BGTH network with a strong emphasis on the application of DBS and delineating the dynamics of PD, there are limitations. As outlined in section \ref{method:discrete_model}, a network of fifty neurons following the ODEs in section \ref{method:modified_HH} was simulated for the STN, GPe, GPi, and TH, and the average currents were added to their overall membrane voltages. This was simulated independently for each region. Consequently, this strategy lacks the physiological realism in imitating how neurons across different portions of the brain connect and interact. Hence, the internal, cross-region brain dynamics are not fully rendered, which can affect the accuracy of the results, and makes the closed-loop system seen here unoptimized. A potentially more suitable approach would have been to establish the network connections of the neurons as shown in Fig. \ref{fig:nuclei}. Due to their physiologically accurate linkages in terms of providing excitatory and inhibitory inputs to each other, building the model to reproduce this idea augments the closed-loop modulation which more correctly reflects the self-regulating facets of the brain. \cite{farkhondeh_tale_navi_closed-loop_2022} As tractography data improves and supplementary information regarding the physiological links across brain regions emerges, a more united network that expands on the technique utilized in this study may yield further information on the behaviour of the brain and the pathology of PD.

\section{\label{conclusion}Conclusion and Future Directions}
The goal of our work was to explicate the elaborate processes that constitute neural functioning, specifically examining regions within the BGTH and contrasting the healthy and Parkinsonian state. To achieve this, we created a co-simulation multiscale model that considers the intricate mechanisms of neuronal firing, diffusion, presynaptic inputs, and the effects of DBS on the brain network. Through this application, we were able to record the dynamics at both the micro- and macroscale, more closely capturing the complexities of the brain. Our findings suggest that irregular thalamic spiking may naturally occur in the context of the connections and conductance changes in the network, rather than being provoked purely by noise. Despite this, noise still has a notable influence on voltage patterns and timing, mimicking the realistic behaviour of neurons.

Regardless of these insights, many uncertainties revolving around the neural dynamics --- especially under the circumstances of PD and other NDDs --- still remain. Thus, our model can serve as a stepping stone to future work that can expand on the findings described here. One direction that should be investigated is the inclusion of closed-loop DBS which can contribute to the continued development of adaptive DBS. Adaptive DBS employs a closed-loop framework that can regulate the stimulation parameters based on feedback received by the patient's brain. By catering the output to the state of each individual, this increases the treatment’s effectiveness and potentially reduces the risk of adverse effects. \cite{krauss_technology_2021} Another beneficial scope to examine is introducing other aspects of the brain into these multiscale models. By expanding to include other vital regions, the prospect of further illuminating the underlying processes of the brain and its involvement in PD is facilitated. Lastly, in our integrated brain connectome data, only one node is associated with each region (STN, GPe, GPi, and TH), and each one is connected with others as presented in Fig. \ref{fig:nodalnetwork}. Therefore, it would be interesting to study when high-resolution brain connectome data is available, where each region consists of more than one node, so that the internal structural organization within a region can also be represented. Such a framework would allow us to capture finer details of intra-regional interactions, rather than reducing an entire area of the brain to a single representative point. Furthermore, considering multiple nodes per sector would open the possibility of exploring heterogeneity within the same sector, such as differences in excitability, connectivity strength, or functional roles of neuronal populations. This could lead to richer dynamical models that better approximate realistic brain activity. Ultimately, such high-resolution connectome studies would provide valuable insights into how both local and global interactions contribute to emergent cognitive processes and neurological disorders. Understanding the intricacies of the brain has been, and still, presents itself to be an immense hurdle. However, by persisting in refining our methods, we strive to be able to exercise our findings to have therapeutical application to those affected by PD, and eventually, all NDDs. 

\begin{acknowledgments}
The authors appreciate the support provided to Prof. Hina Shaheen from the start-up grant at the University of Manitoba and the NSERC Undergraduate Student Research Awards (USRA) program. The authors are grateful to Dr. Swadesh Pal of Wilfrid Laurier University for his insightful comments on the initial version of the manuscript. 
\end{acknowledgments}

\section*{Data availability}
The datasets generated and/or analyzed during the current study are available in the Human Connectome Project repository (https://braingraph.org/cms/).

\section*{References}
\bibliography{references}

\end{document}